\documentclass[aip,jap,reprint]{revtex4-2}

\usepackage{hyperref}
\usepackage{graphicx}
\usepackage{upgreek}
\usepackage{amsmath}
\usepackage{mathtools}
\usepackage{xcolor}

\let\savecorresponds\corresponds
\let\corresponds\relax
\usepackage{mathabx}
\let\corresponds\savecorresponds

\def\eu{\ensuremath{\mathrm{e}}}

\def\du{\ensuremath{\mathrm{d}}}

\def\IL{\ensuremath{I}}
\def\IT{\ensuremath{\tilde{I}}}
\def\PSF{\ensuremath{p}}
\def\Es{\ensuremath{{\cal E}}}
\def\BL{\ensuremath{B}}
\def\BT{\ensuremath{\tilde{B}}}
\def\conv{\ensuremath{\oasterisk}}
\def\deconv{\ensuremath{\odiv}}

\hypersetup{
	colorlinks,
	citecolor=black,
	filecolor=black,
	linkcolor=black,
	urlcolor=black
}

\begin{document}

\title{Aberration control in quantitative widefield quantum microscopy}

\author{S. C. Scholten}
\affiliation{School of Physics, University of Melbourne, VIC 3010, Australia}
\affiliation{Centre for Quantum Computation and Communication Technology, School of Physics, University of Melbourne, VIC 3010, Australia}

\author{I. O. Robertson}
\affiliation{School of Science, RMIT University, Melbourne VIC 3001, Australia}

\author{G. J. Abrahams}
\affiliation{School of Science, RMIT University, Melbourne VIC 3001, Australia}

\author{Priya Singh}
\affiliation{School of Science, RMIT University, Melbourne VIC 3001, Australia}

\author{A. J. Healey}
\affiliation{School of Physics, University of Melbourne, VIC 3010, Australia}
\affiliation{Centre for Quantum Computation and Communication Technology, School of Physics, University of Melbourne, VIC 3010, Australia}

\author{J.-P. Tetienne} 
\email{jean-philippe.tetienne@rmit.edu.au}
\affiliation{School of Science, RMIT University, Melbourne VIC 3001, Australia}

\date{\today}

\begin{abstract}
Widefield quantum microscopy based on nitrogen-vacancy (NV) centres in diamond has emerged as a powerful technique for quantitative mapping of magnetic fields with a sub-micron resolution. 
However, the accuracy of the technique has not been characterised in detail so far.
Here we show that optical aberrations in the imaging system may cause large systematic errors in the measured quantity beyond trivial blurring.
We introduce a simple theoretical framework to model these effects, which extends the concept of a point spread function to the domain of spectral imaging. Using this model, the magnetic field imaging of test magnetic samples is simulated under various scenarios, and the resulting errors quantified.  
We then apply the model to previously published data, show that apparent magnetic anomalies can be explained by the presence of optical aberrations, and demonstrate a post-processing technique to retrieve the source quantity with improved accuracy.
This work presents a guide to predict and mitigate aberration induced artefacts in quantitative NV-based widefield imaging and in spectral imaging more generally.
\end{abstract}

\maketitle

\section{Introduction}

A broad class of quantum sensors relies on the optical readout of a property of an atomic or atom-like system, such as the transition frequency between two discrete energy levels or the quantum phase of a superposition of states.\cite{degenQuantumSensing2017} 
A prime example of such a quantum sensor is the nitrogen-vacancy (NV) centre in diamond, whose electron spin can be readout optically allowing the measurement of its spin resonance, spin lifetime and spin coherence. 
These quantities in turn can be used to extract quantitative information about the local environment, such as the static magnetic or electric field, or the spectral density of magnetic fluctuations.\cite{rondinMagnetometryNitrogenvacancyDefects2014,casolaProbingCondensedMatter2018} 
A strength of these quantum sensors is their accuracy, which derives from the fact that the quantum property being measured has a well defined relationship with the quantity of interest. 
For instance, under appropriate conditions the frequency splitting in the electron spin resonance spectrum of the NV centre is given by $\Delta\omega=2\gamma_{\rm NV}B_{\rm NV}$ where $\gamma_{\rm NV}/2\pi=28.033(3)$\,GHz\,T$^{-1}$ is the NV gyromagnetic ratio and $B_{\rm NV}$ is the magnetic field projection along the NV symmetry axis.\cite{dohertyNitrogenvacancyColourCentre2013,rondinMagnetometryNitrogenvacancyDefects2014} 
As the constant $\gamma_{\rm NV}$ has been precisely measured and is expected to be robust against a wide range of perturbations,\cite{dohertyNitrogenvacancyColourCentre2013,feltonHyperfineInteractionGround2009} the NV centre is commonly used as an absolute, calibration-free magnetometer (measuring the projection $B_{\rm NV}$).         

However, when quantum sensors are employed to image spatial variations of a measurand, additional complexities in the readout system may cause systematic errors in the apparent value of the measurand across the image. 
In particular, widefield optical readout of an array of quantum sensors is a convenient and rapid way to perform imaging with a sub-micrometer resolution (limited by optical diffraction) over a large field of view.\cite{steinertHighSensitivityMagnetic2010,levinePrinciplesTechniquesQuantum2019,scholtenWidefieldQuantumMicroscopy2021}
However, the widefield platform's parallel nature makes it prone to cross-talks due to optical aberrations, raising the question of the accuracy of the supposedly local measurement. 
Here we define aberration as the departure from an ideal imaging system -- which maps a point in the object plane (the array of quantum sensors) onto a single point in the image plane (the camera sensor) -- causing a spread of the light in the image plane, characterised by the point spread function (PSF).\cite{braatAssessmentOpticalSystems2008}
Sources of aberrations may include optical diffraction, the non-ideal focusing properties of the lenses employed, light scattering along the imaging path, and experimental imperfections such as defocusing and lateral drifts during data collection. 
In this paper, we analyse the impact of optical aberrations on the accuracy of widefield quantum microscopy, by introducing a simple model that extends the concept of PSF to the case of spectral imaging.
Specifically, we consider the type of spectral imaging where the quantity of interest resides in the spectral peak positions, as is the case with NV-based static magnetic imaging.
We note, however, that our model can be easily extended to other forms of spectral imaging including outside quantum microscopy, e.g.\ fluorescence lifetime imaging.\cite{steinertMagneticSpinImaging2013,simpsonElectronParamagneticResonance2017,webbWidefieldTimedomainFluorescence2002,bowmanElectroopticImagingEnables2019} 

\begin{figure*}[tb!]
	\includegraphics[width=0.85\textwidth]{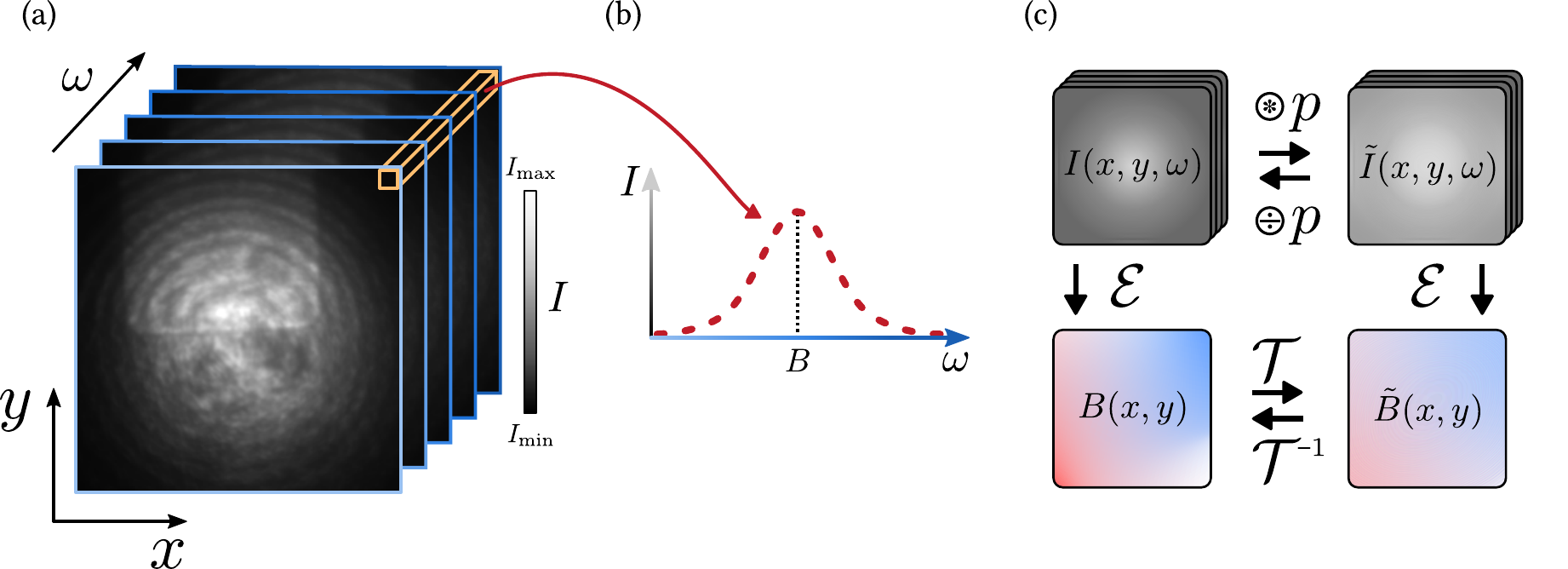}
	\caption{\textbf{Optical aberrations in spectral imaging.} (a)~The data cube $I(x,y,\omega)$ is composed of many images [with spatial dimensions $(x,y)$] corresponding to different values of the spectral variable $\omega$.
	(b)~Spectrum $I(x_i,y_i,\omega)$ corresponding to a given pixel $(x_i,y_i)$, from which a quantity of interest $B(x_i,y_i)$, e.g.\ the peak position, is extracted via an estimator \Es.
	(c)~Optical aberrations transform each ideal optical image $I(x,y,\omega_i)$ into a blurred image $\tilde{I}(x,y,\omega_i)$ through a convolution with the point spread function \PSF.
	As a result, the apparent map $\tilde{B}(x,y)$ is a distorted version (transformation ${\cal T}$) of the ideal map $B(x,y)$ that would be obtained in the absence of aberrations.}
	\label{fig:principle}
\end{figure*}

Widefield imaging of arrays of NV centres in diamond, also known as quantum diamond microscopy,\cite{levinePrinciplesTechniquesQuantum2019,scholtenWidefieldQuantumMicroscopy2021} has been increasingly used for the rapid, quantitative mapping of magnetic fields and other quantities (electric field\cite{broadwaySpatialMappingBand2018}, strain\cite{broadwayMicroscopicImagingStress2019,kehayiasImagingCrystalStress2019}), especially for the characterisation of external samples such as magnetic materials\cite{lesageOpticalMagneticImaging2013,glennMicrometerScaleMagnetic2017,torailleOpticalMagnetometrySingle2018,fescenkoDiamondMagneticMicroscopy2019,broadwayImagingDomainReversal2020,meirzadaLongtimescaleMagnetizationOrdering2021,mclaughlinStrongCorrelationSuperconductivity2021} or current-carrying electronic devices\cite{nowodzinskiNitrogenvacancyCentersDiamond2015,tetienneQuantumImagingCurrent2017,kuImagingViscousFlow2020,turnerMagneticFieldFingerprinting2020,lillieImagingGrapheneFieldeffect2019,scholtenImagingCurrentPaths2022} placed in close proximity to the diamond. 
For these applications, the accuracy of the magnetic field measurements is critical as it is used to estimate materials parameters or analyse transport phenomena in detail. 
In particular, our study is motivated by two recent observations.
First, a study by Tetienne et al.\ has uncovered an apparent anomaly in the magnetic field generated by current-carrying metallic wires fabricated on the diamond sensor.\cite{tetienneApparentDelocalizationCurrent2019} 
This apparent anomaly, as we will show, can be resolved via the consideration of optical aberrations. 
Second, Fu et al.\ have observed the presence of background NV photoluminescence (PL) in a quantitative study of magnetic rock samples, and applied a correction to the apparent magnetic field (they multiplied by a factor of 2) to account for this background contribution.\cite{fuHighsensitivityMomentMagnetometry2020} 
We will present a simple model to systematically predict the transformation required to correct aberrant raw data in a range of scenarios. 

The paper is organised as follows. 
We first present the general theoretical framework for accounting for optical aberrations in spectral imaging, and specify it to the case of spectral peak finding (Sec. \ref{sec:theory}).
We then use the model to simulate the magnetic field imaging of two test systems, a magnetic material and a current-carrying wire, under various scenarios of optical aberrations and illumination conditions (Sec. \ref{sec:sims}).   
The resulting errors and imaging artefacts will be discussed qualitatively and characterised quantitatively.
We will then demonstrate a post-processing technique utilizing basic deconvolution algorithms to retrieve the source quantity with improved accuracy (Sec.~\ref{sec:deconv}).
Finally, we will apply the model to experimental data reported by Tetienne et al.\cite{tetienneApparentDelocalizationCurrent2019} and resolve the apparent magnetic anomaly (Sec. \ref{sec:anom}).
% Finally, we will apply the model to resolve the apparent magnetic anomaly reported by Tetienne et al.,\cite{tetienneApparentDelocalizationCurrent2019} and will demonstrate a post-processing technique to retrieve the source quantity with improved accuracy. 
This work will serve as a guide to correctly analyse data from quantitative widefield imaging based on NV centres or similar quantum sensors. It may also be relevant to similar peak finding problems in spectral imaging outside quantum microscopy, and can be extended to other forms of spectral imaging such as lifetime imaging.

\section{Theoretical framework}
\label{sec:theory}

\subsection{General problem}

The principle of spectral imaging is depicted in Fig.~\ref{fig:principle}(a). 
In the case of widefield imaging, the data cube $I(x,y,\omega)$ is acquired by optically recording images [spatial dimensions $(x,y)$] for various values of the spectral variable $\omega$.
A spectrum $I(x_i,y_i,\omega)$ can then be formed at every pixel $(x_i,y_i)$ and analysed to extract an estimate of a quantity of interest, $B(x_i,y_i)=\Es[I(x_i,y_i,\omega)]$ where $\Es$ is the estimator [Fig.~\ref{fig:principle}(b)].

The optical response of the imaging system is modelled using the concept of a PSF, defined as the system's impulse response $\PSF(x,y)$. 
If $\IL(x,y,\omega_i)$ is the true optical intensity map of the object, for instance the PL emission map of an array of NV centres, then the intensity map recorded by the imaging system for each $\omega_i$ is a convolution with the PSF,
\begin{align}\label{eq:pl_psf}
\begin{split}
    \IT(x,y,\omega_i) &= \iint \IL(x', y',\omega_i) p(x - x', y - y')\ \du{x'}\du{y'}
    \\
    &= \{\IL \conv p\}(x,y,\omega_i)\,, % useful?
\end{split}
\end{align}
where $\conv$ is the convolution operator. 
We will use $\IL$ and $\IT$ to refer to true and apparent intensity maps, respectively. 
Note that throughout the paper continuous integrals are used for generality rather than the discrete sums required in practice.

One approach to account for the optical response of the imaging system is to deconvolve Eq.\,\ref{eq:pl_psf} for $\IL$ at each $\omega$ value (i.e.\ correcting the whole data cube), before attempting to extract $\BL$. 
However, this process is not only computationally intensive, but the PSF is often a priori unknown.
In widefield NV experiments the PSF may be especially difficult to measure as optical aberrations will vary between samples and cannot be easily characterised in situ.
Therefore, it is important to develop an understanding of how a non-ideal PSF affects the estimate based on the uncorrected (blurred) data, and develop strategies to efficiently predict and correct errors.  

When the experimentally determined data cube $\IT(x,y,\omega)$ is analysed by applying the estimator $\Es$ pixel by pixel, a spatial map of the estimate is obtained,
\begin{align}\label{eq:BT}
\begin{split}
    \BT(x,y) &= \Es[\IT(x,y,\omega)]
    \\
    &= \Es\left[\iint \IL(x', y',\omega)p(x - x', y - y')\ \du{x'}\du{y'}\right]
    \\
    &= {\cal T}[\BL(x,y)]
\end{split}
\end{align}
where ${\cal T}$ is the forward transformation we seek to determine [Fig.~\ref{fig:principle}(c)]. 
In general, this transformation from the true estimate $\BL$ to the apparent estimate $\BT$ cannot be cast in terms of a convolution with a PSF -- not even a modified PSF.

The goal of this paper is to examine this transformation in scenarios relevant to quantum microscopy. 
Specifically, we will consider the important case of peak finding, when the estimator is chosen to determine the position of a peak in the spectrum. 
This position may be, for instance, the resonance frequency of an electron spin transition of the NV centre in the magnetic resonance spectrum, which is then converted e.g.\ to a magnetic field (often after two or more resonance frequencies are determined, and using the expression for the Zeeman splitting $\Delta\omega=2\gamma_{\rm NV}B_{\rm NV}$). 
However, we stress that the presented framework can be applied (at least numerically) to analyse any estimator, including any arbitrary data fitting algorithm. 
For example, when the spectral variable corresponds to a time delay ($\omega\triangleq\tau$) and the optical intensity decays over time according to $\IL(\tau)=\IL(0)\eu^{-\tau/\tau_c}$, a common quantity of interest is the characteristic time of the decay, $\tau_c$. 
In this case, a common estimator is the value $\tau_c$ of the best fit to the data using the assumed exponential decay function. 
This scenario applies, for instance, to fluorescence lifetime imaging,\cite{webbWidefieldTimedomainFluorescence2002,bowmanElectroopticImagingEnables2019} or with NV centres to spin lifetime (relaxometry) and decoherence imaging.\cite{steinertMagneticSpinImaging2013,simpsonElectronParamagneticResonance2017} 

\subsection{Peak finding}

We now derive an explicit expression for the transformation ${\cal T}$ in the case where the quantity of interest is the position of a peak in the spectrum. 
While $B$ is typically estimated through fitting the spectrum to an appropriate model (e.g.\ a Gaussian or Lorentzian function), for our purpose it is convenient to consider the simplest possible estimator suitable for peak finding, namely we define $B$ as the expectation value of the variable $\omega$ with probability distribution $I(\omega)$,   
\begin{align}\label{eq:Bmean}
    \BL(x,y) &= \frac{\int \omega \IL(x,y,\omega) \du{\omega}}{\hphantom{\omega}\int \IL(x,y,\omega) \du{\omega}}.
\end{align}
In the case of a spectrum with zero baseline and a symmetric lineshape as depicted in Fig.~\ref{fig:principle}(b), this estimator yields the central position of the peak. 
Thus, it constitutes a good model of peak fitting algorithms that are typically used to analyse more complex spectra (e.g.\ with a non-zero or even non-constant baseline, or multiple partially overlapping peaks).   

Applying the estimator Eq.~\ref{eq:Bmean} to the experimentally determined data cube $\IT(x,y,\omega)$, Eq.~\ref{eq:BT} becomes 
\begin{align}\label{eq:B_psf}
\begin{split}
    &\BT(x,y) = \frac{\int \omega \IT(x,y,\omega)\  \du{\omega}}{\int \IT(x,y,\omega)\ \du{\omega}}
    \\
    &= \frac{\int \omega \iint \IL(x', y',\omega) p(x - x', y - y')\ \du{x'}\du{y'}\du{\omega}}{\int \IT(x,y,\omega)\ \du{\omega}}
    \\
    &= \frac{\iint \BL(x', y',\omega) \int \IL(x',y',\omega)\  \du{\omega} \,p(x - x', y - y')\ \du{x'}\du{y'}}{\int \IT(x,y,\omega)\ \du{\omega}}.
\end{split}
\end{align}
We can further simplify the expression by introducing the mean optical intensity across the spectrum, 
\begin{align}\label{eq:Im}
 \IL_m(x,y)=\frac{1}{\omega_{\rm max}-\omega_{\rm min}}\int \IL(x,y,\omega)\ \du{\omega}
\end{align}
where $\omega_{\rm min,max}$ are the bounds of the integral, with a similar definition for the mean intensity of the apparent spectrum, $\IT_m(x,y)$. 
This definition gives
\begin{align}\label{eq:B_psf_final}
\begin{split}
    \BT(x,y) &= \iint \BL(x', y')p(x - x', y - y')\frac{\IL_m(x',y')}{ \IT_m(x\hphantom{'},y\hphantom{'})}\ \du{x'}\du{y'}.
\end{split}
\end{align}

As we can see, in general the apparent $\BT(x,y)$ map is not simply a convolution of $\BL(x,y)$ with the optical system's PSF, because of the factor $\IL_m(x',y') /  \IT_m(x,y)$.
However, we can re-write Eq.~\ref{eq:B_psf_final} as a convolution of the product $\BL\IL_m$,
\begin{align}\label{eq:BIconv}
\begin{split}
    \BT\IT_m(x,y) &= \{\BL\IL_m \conv p\}(x,y)
\end{split}
\end{align}
such that the true estimate $\BL$ can be recovered with a simple operation on the raw data,
\begin{align}\label{eq:Bdeconv}
\begin{split}
    \BL &= \frac{\{\BT\IT_m \deconv p\}}{\{\IT_m \deconv p\}} \\
    &= {\cal T}^{-1}[\BT]\,
\end{split}
\end{align}
where $\deconv$ represents a deconvolution operation.
Equation~\ref{eq:Bdeconv} expresses the inverse transformation ${\cal T}^{-1}$ to be applied to the normally processed data to account for optical aberrations [Fig.~\ref{fig:principle}(c)].
This procedure is vastly simplified compared with the brute force approach of deconvolving the entire data cube before processing (2 deconvolutions instead of hundreds typically).  
We will now examine a few limit cases for the PSF $p$ and the mean intensity map $I_m$. 

\subsection{Limit cases}

First, if the optical imaging system is ideal, i.e.\ $\PSF(x,y)=\delta(x,y)$ where $\delta$ is the Dirac delta function, then the apparent estimate obviously matches the true one at every pixel, $\BT(x,y)=\BL(x,y)$. 

In practice, the PSF always has a finite width corresponding to at least the optical diffraction limit. 
An interesting limit case is then found when the optical intensity is completely uniform across the field of view, $\IL_m(x,y)=\IT_m(x,y)=I_0$. 
In this case, Eq.~\ref{eq:B_psf_final} becomes a trivial convolution, $\BT=\{\BL \conv p\}$. 
That is, the apparent estimate $\BT$ is simply a blurred version of the true estimate $\BL$ in the same way every individual optical image is blurred.

A particular model of a PSF relevant to widefield NV experiments deserves special consideration. 
Indeed, it has been observed that the PL image of an NV layer as recorded by a camera contains not only an image of the object but also a background contribution.\cite{fuHighsensitivityMomentMagnetometry2020} 
This background is believed to be the result of reflections of the light emitted by the NVs on the surfaces of the diamond slab, forming secondary defocused images on the camera sensor.\cite{phamMagneticFieldImaging2011}
More generally, any form of light scattering will add a background contribution if it reaches the camera sensor. 
This background light still contains spectral information but is now delocalised across many pixels of the recorded image. In the case of a full delocalisation on the scale of the camera field of view, we can write the PSF as
\begin{align}\label{eq:psf_cste}
\begin{split}
    p(x,y) &= (1-\alpha_b)\delta(x,y)+\alpha_b/A
\end{split}
\end{align}
where the primary image is modelled as an ideal (delta function) response for simplicity, and $\alpha_b$ is the fraction of background contribution ($A$ is the area of the field of view, ensuring that $p$ is normalised to unity). We then obtain 
\begin{eqnarray}\label{eq:b_app_psf_cste}
    \BT(x,y) = (1-\alpha_b)\BL(x,y)+\alpha_b B_m
\end{eqnarray}
where 
\begin{eqnarray}\label{eq:b_app_mean}
    B_m = \frac{1}{A}\iint \BL(x,y)\ \du{x}\du{y}
\end{eqnarray}
is the mean value of the estimate over the field of view.
In imaging experiments, we are often interested in measuring small local changes of a quantity $B$ relative to a uniform background value $B_m$. 
This difference $\Delta B=B-B_m$ is then
\begin{eqnarray}\label{eq:b_diff}
    (\BT(x,y)-B_m) &=& (1-\alpha_b)(\BL(x,y)-B_m) \\
    \Delta\BT(x,y) &=& (1-\alpha_b)\Delta\BL(x,y).
\end{eqnarray}
Thus, any local change $\Delta\BL$ will appear reduced by the fraction $\alpha_b$. 
In this scenario, the correction to apply to the processed data $\Delta\BT$ is trivial, although precise determination of the fraction $\alpha_b$ is critical to the resulting accuracy. 
In the magnetic imaging experiments of Fu et al.,\cite{fuHighsensitivityMomentMagnetometry2020} the authors estimated a background contribution of $\alpha_b\approx0.5$ by an independent optical characterisation, and corrected their apparent magnetic field maps by inverting Eq.~\ref{eq:b_diff}, i.e.\ $\Delta\BL=\Delta\BT/(1-\alpha_b)\approx2\Delta\BT$.      

In general, however, the PSF will likely be less trivial than the simple case examined above.
Moreover, the uniform intensity assumption is rarely valid because of spatial inhomogeneities in the emitting object or in the illumination conditions. 
For instance, in widefield NV experiments the laser illumination spot has a finite size and suffer from inhomogeneities (including due to interference effects), and the light collected from the NV layer is also not uniform either due to intrinsic spatial variations in NV properties or density, or due to the presence of a sample external to the diamond that modulates the optical emission or collection. 
In Sec.\,\ref{sec:sims}, we numerically simulate some of these scenarios.

\section{Simulating aberrant magnetic images} \label{sec:sims}

\begin{figure*}[tb]
	\includegraphics[width=0.99\textwidth]{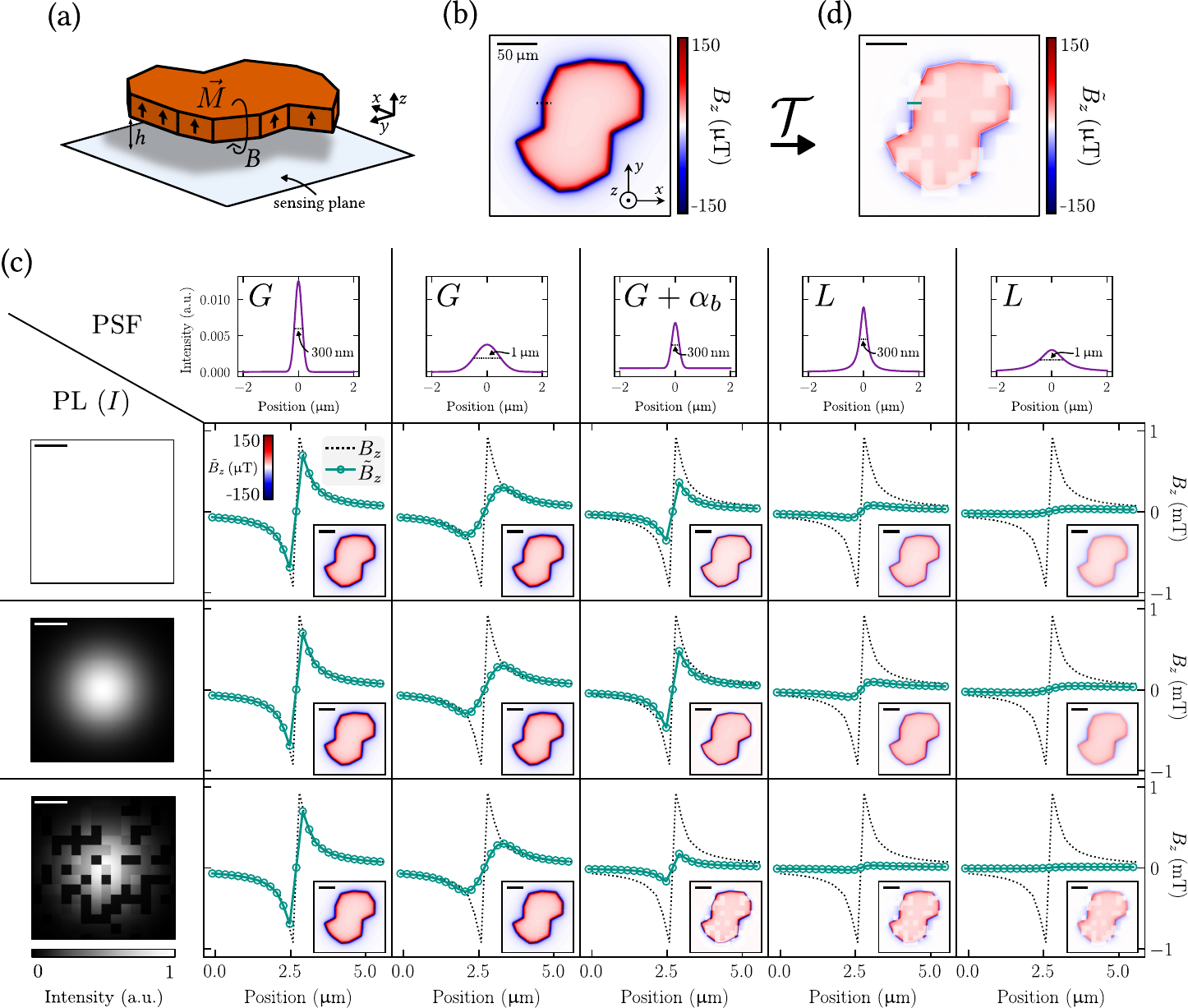}
	\caption{\textbf{Optical aberrations in magnetic imaging of ferromagnetic materials.}
	(a)~A micromagnet creates a stray magnetic field in the sensing plane at a standoff distance $h$.
	(b)~Theoretical magnetic field map (out-of-plane component $B_z$) for a perpendicularly-magnetized flake with magnetisation $M_z=1$~MA/m, thickness $t=1$~nm and standoff $h=100$~nm.
	(c)~Table showing the apparent magnetic field image ($\BT_z$) and a line cut through an edge of the flake [along black dotted line in (b)] under various scenarios. 
	The 3 rows of the table correspond to different true optical intensity patterns (i.e.\ distribution $I$, shown on the far left): uniform illumination (top row), Gaussian illumination with FWHM\,$=100\,\upmu$m (middle), Gaussian illumination modulated by a random chessboard pattern at relative intensities of $0.1$, $0.8$ and $1.0$ (bottom). 
	The 5 columns of the table corresponds to different axially symmetric PSFs ($p$, line cut shown on the far top): Gaussian ($G$) with FWHM\,$=300$\,nm (column 1), Gaussian with FWHM\,$=1\,\upmu$m (column 2), Gaussian with FWHM\,$=300$\,nm and a constant offset $\alpha_b=0.5$ (column 3), Lorentzian ($L$) with FWHM\,$=300$\,nm (column 4), and Lorentzian with FWHM\,$=1\,\upmu$m (column 5). 
	(d)~Magnified version of the apparent $\BT_z$ map for the case of Gaussian illumination with chessboard pattern and Lorentzian PSF with FWHM\,$=1\,\upmu$m. The colour scale in all magnetic field maps in (b-d) is capped to $\pm150\,\upmu$T.}
	\label{fig:forward}
\end{figure*}

To illustrate the impact optical aberrations have both qualitatively (imaging artefacts) and quantitatively (systematic errors), we calculate the apparent $\BT$ using Eq.~\ref{eq:B_psf_final} under various conditions for two typical test problems: imaging of the stray magnetic field from a micromagnet, and imaging of the \O rsted magnetic field from an electric current.

\subsection{Stray field imaging of a micromagnet}
\label{subsec:mag_sim}

We first simulate magnetic field imaging of a micromagnet, as would be performed using arrays of NV centres.\cite{lesageOpticalMagneticImaging2013,glennMicrometerScaleMagnetic2017,torailleOpticalMagnetometrySingle2018,fescenkoDiamondMagneticMicroscopy2019,broadwayImagingDomainReversal2020,meirzadaLongtimescaleMagnetizationOrdering2021,mclaughlinStrongCorrelationSuperconductivity2021}
The simulated micromagnet is a flake-like structure (thickness $t=1$\,nm, lateral size $\sim100\,\upmu$m) with uniform out-of-plane magnetisation $M_z=1\,{\rm MA/m}$ (resulting in an areal magnetization $M_zt=1$\,mA) placed near the array of quantum sensors (standoff $h=100$\,nm) [Fig.~\ref{fig:forward}(a)].
The stray magnetic field produced by the flake is measured in the plane of the sensor, projected onto a given axis; here we consider the $z$ projection, denoted as $\BL_z$. The true magnetic field map $\BL_z(x,y)$ for the simulated flake is shown in Fig.~\ref{fig:forward}(b).

We now compute via Eq.~\ref{eq:B_psf_final} the apparent field map, $\BT_z(x,y)$, for various assumptions for the PSF $p$ and the mean optical intensity map $I$ [previously denoted as $I_m$, but for simplicity we will drop the $m$ subscript in the rest of the paper]. 
The results are presented in Fig.~\ref{fig:forward}(c) in the form of a table, where each column corresponds to a different PSF, and each row corresponds to a different intensity pattern. 
Each element in the table shows the resultant apparent field map $\BT_z(x,y)$, and a line cut comparison with the ground truth $\BL_z$, for the associated PSF and intensity pattern.  

The optical intensity map recorded by the camera sensor is the result of the illumination pattern (how much light the quantum sensors receive in the first place), the optical response of the quantum sensors (how much light they re-emit in the form of PL), and the collection efficiency (how much of the emitted PL makes it to the camera sensor). 
The measured intensity map will therefore depend on many factors including the profile of the excitation beam, the homogeneity of the quantum sensing array, and the optical properties of the magnetic sample. 
The latter is particularly important with metallic samples, which may locally quench the PL emission or enhance the PL collection depending on the standoff distance.\cite{tetienneApparentDelocalizationCurrent2019} 
Here we consider three distinct patterns displayed in the far left column of Fig.~\ref{fig:forward}(c): (i)~a spatially uniform intensity (top row); (ii)~a Gaussian profile, at roughly the width of the sample (100\,$\upmu$m) (middle row); and (iii)~the same Gaussian profile modulated with a random chessboard pattern at relative intensity of 0.1, 0.8 and 1.0 (bottom row).
Case (i) represents the ideal imaging situation, (ii) the more common situation where the sensing array is illuminated by a finite-size but smooth laser beam, and (iii) a worst-case scenario where the optical intensity varies on a scale comparable to that of the magnetic field variations, due e.g.\ to laser beam imperfections, inhomogeneities in the sensing array, or quenching by surface contaminants.
For the PSF, we model a few important examples, displayed in the far top row of Fig.~\ref{fig:forward}(c), namely Gaussian and Lorentzian profiles with full width at half maximum (FWHM) of 300\,nm or $1\,\upmu$m, as well as a Gaussian profile with a vertical offset similar to the limit case expressed by Eq.~\ref{eq:psf_cste}.
The 300\,nm Gaussian PSF roughly corresponds to the diffraction limit with a high numerical aperture (NA) objective lens.
The Lorentzian and wider Gaussian profiles thus represent different scenarios of aberrations beyond the diffraction limit.

Examining the results in Fig.~\ref{fig:forward}(c), we see that a near-diffraction-limited Gaussian response (column 1) produces an apparent field quantitatively very similar to the true field, with only a very minor blurring, regardless of the optical intensity pattern.
A wider Gaussian PSF response (column 2) shows more blurring, but no anomalous quantitative features for any of the intensity patterns.
Adding a global offset to the optical signal of $\alpha_b=0.5$ to the PSF of column 1 (column 3) produces a field with sharp features (i.e.\ no additional blurring) of a reduced magnitude.
Crucially, the field magnitude reduction depends on the optical intensity pattern. 
For instance, the Gaussian illumination leads to a smaller apparent reduction compared to uniform illumination (31\% vs 48\% reduction, relative to the $\alpha_b=0$ case).
We see this change because in the uniform illumination case there is a larger contribution from regions far from the flake which experience no magnetic field. 
Finite-size (Gaussian) illumination helps to restrict those long-range contributions but also leads to more subtle errors, for instance the apparent field becomes slightly positive outside the flake (as evidenced by the light red background in the image) when the true field has a negative tail there; this is because the apparent field outside the flake where the true PL is low is dominated by contributions from inside the flake where the PL is high. 
We note that $\alpha_b=0.5$ means that 50\% of the PL collected from a given quantum sensor is delocalised across the entire field of view, which models the situation in Ref.~\onlinecite{fuHighsensitivityMomentMagnetometry2020}, highlighting the importance of precise characterisation of both the PSF and optical intensity pattern.
The effect of the global offset on the PSF is more visually apparent for the chessboard PL pattern, where local dark PL regions appear blocky in the field map with a reduced magnitude. 
Again, this difference is because the weight of the local contribution (carrying information about the true local field) is being modulated, locally changing the apparent net field. 
The Lorentzian PSF (columns 4 and 5) leads to an apparent magnetic field that displays the blocky anomalous measurement of column 3 for a chessboard PL pattern, see Fig.~\ref{fig:forward}(d) for a magnified image. 
Additionally, a significant blurring and reduction in apparent field strength are observed for each PL pattern.
This reduction is seen even with FWHM\,$=300$\,nm due to the wide tails of the Lorentzian profile which dramatically increase the effect of remote contributions. 
Taking the uniform PL case, we find an apparent reduction in maximum field strength (relative to the true field) of 91\% for the 300-nm Lorentzian PSF, compared to only 25\% for the 300-nm Gaussian. 
With the 300-nm Lorentzian PSF, the apparent field is also very sensitive to the PL pattern unlike with the Gaussian PSF, with a reduction of up to 97\% (relative to true field) in the most quenched region of the chessboard PL pattern.
These observations highlight again the importance of precisely accounting for both the PSF and the optical intensity pattern.

We note that the blocky imaging artefacts appear where there is minimal local optical signal (squares at intensity of 0.1 relative to the maximum of 1.0) and thus the tails from neighbouring regions dominate, whereas the areas of moderate relative signal (squares at 0.8) report the field more faithfully.
Aberrant measurements can thus be recognised by features in a measurand at locations of low PL intensity.
In practice, however, rather than a random PL modulation pattern as simulated here, PL modulations are often correlated with the true magnetic field distribution itself (e.g. due to PL quenching from the magnetic sample), which means artefacts may be difficult to visually recognise.
Systematically applying the inverse transformation ${\cal T}^{-1}$ to the apparent field will therefore be an important step for accurate quantitative analysis of widefield measurements (see Sec.~\ref{sec:deconv}).

\subsection{\O rsted field imaging of a charge current}
\label{subsec:curr_sim}

\begin{figure*}[tb]
	\includegraphics[width=0.85\textwidth]{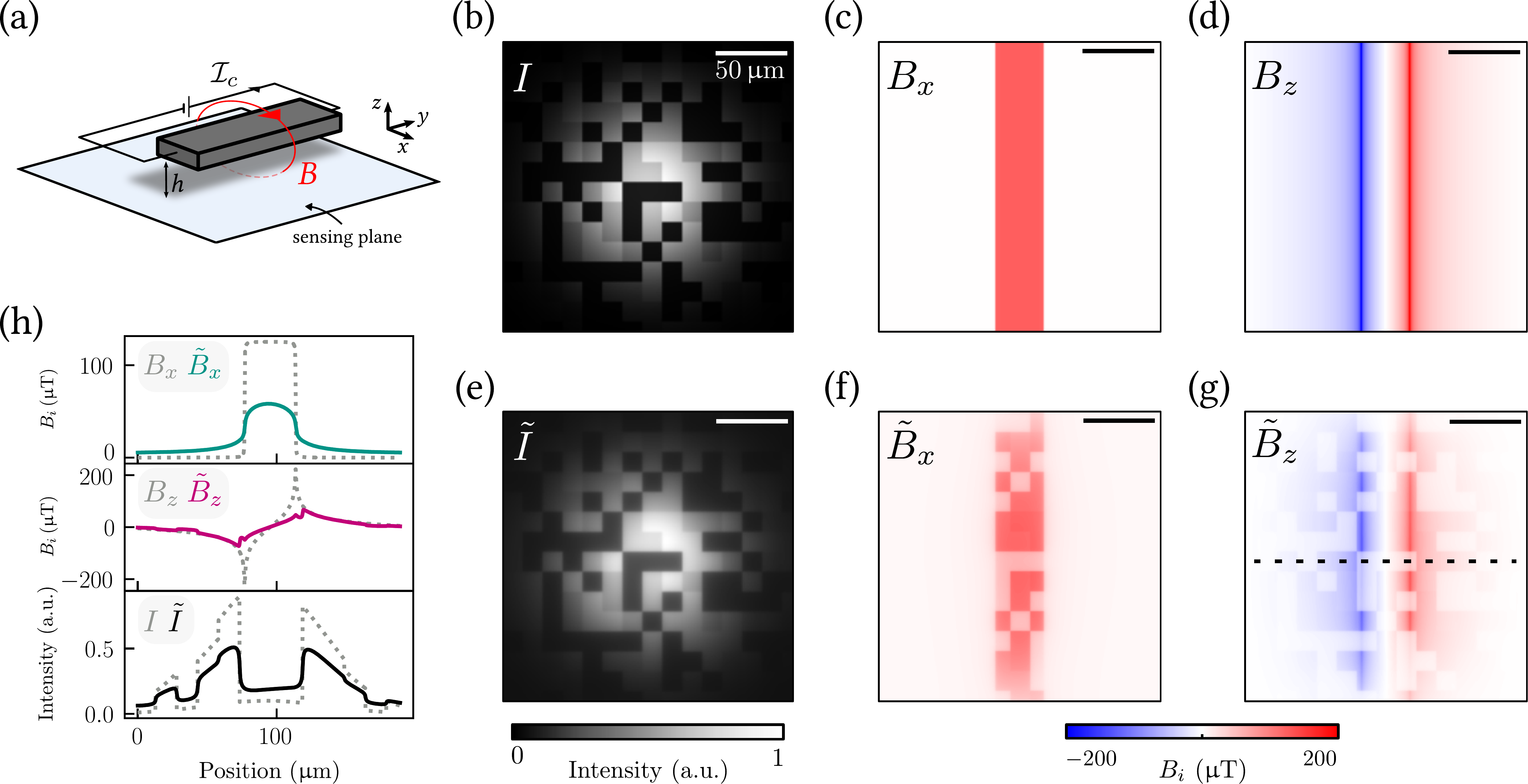}
	\caption{\textbf{Optical aberrations in magnetic imaging of currents.} 
	(a)~A current ${\cal I}_c$ passing in a flat conducting wire (strip) creates an \O rsted magnetic field in the sensing plane at a standoff $h$. 
	(b)~Assumed optical intensity map without aberrations, $\IL$. 
	(c,d)~True magnetic field maps $B_x$ (c) and $B_z$ (d) in the sensing plane (standoff $h=100$\,nm).
	(e)~Apparent optical intensity map ($\IT$) assuming Gaussian illumination with chessboard pattern and Lorentzian PSF with FWHM\,$=1\,\upmu$m. 
	(f,g)~Apparent $\tilde{B}_x$ and $\tilde{B}_z$ maps computed by applying the transformation ${\cal T}$ (Eq.~\ref{eq:B_psf_final}) using (b-e) as inputs. 
	(h)~Line cuts across the wire [dotted line in (g)] of the $B_x$ component (top graph), $B_z$ component (middle) and PL (bottom) comparing true (dotted lines) and apparent quantities (solid lines).}
	\label{fig:current}
\end{figure*}

Magnetic field imaging is also useful to analyse current distributions via the \O rsted magnetic field they produce.\cite{kuImagingViscousFlow2020, lillieImagingGrapheneFieldeffect2019, tetienneQuantumImagingCurrent2017, turnerMagneticFieldFingerprinting2020}
In the context of imaging artefacts, a difference with the case studied above is that the field from a current is typically not confined within the field of view and so edge effects become important, whereas when imaging a magnetic material one often adjusts the field of view in order to fully capture the stray field.    

To simulate this scenario, we consider the current in a flat wire (width $40~\upmu$m) parallel to the diamond surface (standoff $h=100$\,nm) carrying a current density of ${\cal J}_c = 200\,$A/m [Fig.~\ref{fig:current}(a)].
The assumed PL intensity pattern is again a Gaussian-chessboard [Fig.~\ref{fig:current}(b)], and we focus on the $x$ and $z$ components of the \O rsted field. The true field maps $B_x$ and $B_z$ are shown in Fig.~\ref{fig:current}(c,d).
The worst-case PSF from the previous section, a $1\,\upmu {\rm m}$ wide (FWHM) Lorentzian, is utilized to calculate the apparent PL intensity and field [Fig.~\ref{fig:current}(e-g)] according to Eq.~\ref{eq:B_psf_final}.
Both apparent field components have strong suppression of the local field where there is less local optical signal, as can be quantified by the profiles in Fig.~\ref{fig:current}(h). 
Clearly, the apparent field is not simply a blurred version of the true field, for instance $\BT_x$ is reduced by 53\% in the middle of the wire in this region of reduced PL despite the true $B_x$ being uniform over the 40-$\upmu$m width of the wire; highlighting again the far-reaching effect of the Lorentzian tails from neighbouring regions.

Near the borders of the image, the PL intensity is very small due to the Gaussian illumination spot. 
As a result, the apparent field near the top and bottom borders is reduced compared to the centre of the image, despite the true \O rsted field being uniform along the wire. Such a correlation between apparent field and illumination pattern can serve as an indicator of aberrant measurements.
While this indicator is naturally present when imaging current distributions, for magnetic materials it is possible to move the illumination spot in such a way that the magnetic sample is located in the tail of the illumination spot, revealing more clearly the presence or not of a correlation. 

\section{Controlling for aberration in simulated magnetic images} \label{sec:deconv}

\begin{figure*}[tb]
	\includegraphics[width=0.95 \textwidth]{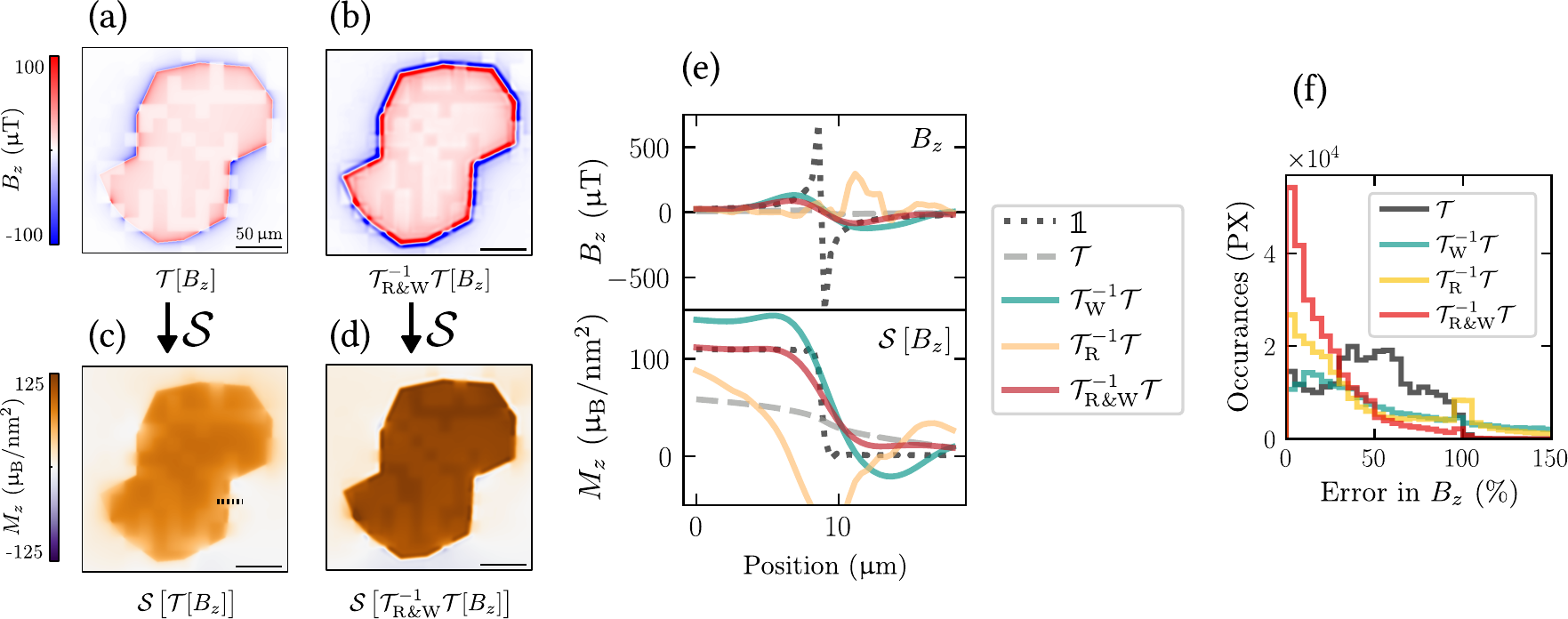}
	\caption{\textbf{Comparing aberration correction methods for micromagnet simulation.}
	(a)~Apparent magnetic field map (out-of-plane component $B_z$) imaged through an optical system with a 1\,$\upmu$m wide (FWHM) Lorentzian PSF, and a chessboard emission pattern, equivalent to Fig.~\ref{fig:forward}(d).
	(b)~True magnetic field image calculated by deconvolving (a) with the R\&W method (see text).
	(c,d)~Source magnetization maps calculated from the field maps in (a,b) respectively. 
	(e)~Line cuts at position indicated in (c) for stray field (top) and magnetization (bottom) images: aberration-free (black dotted), apparent (grey dashed), W-deconvolved (green), R-deconvolved (orange), and R\&W-deconvolved (red).
    (f)~Histograms of the relative error in $B_z$ for each of the deconvolution methods as described above, compared with the apparent case (no deconvolution operation; black line).}
	\label{fig:mag_deconv}
\end{figure*}

\begin{figure*}[tb] 
	\includegraphics[width=0.7 \textwidth]{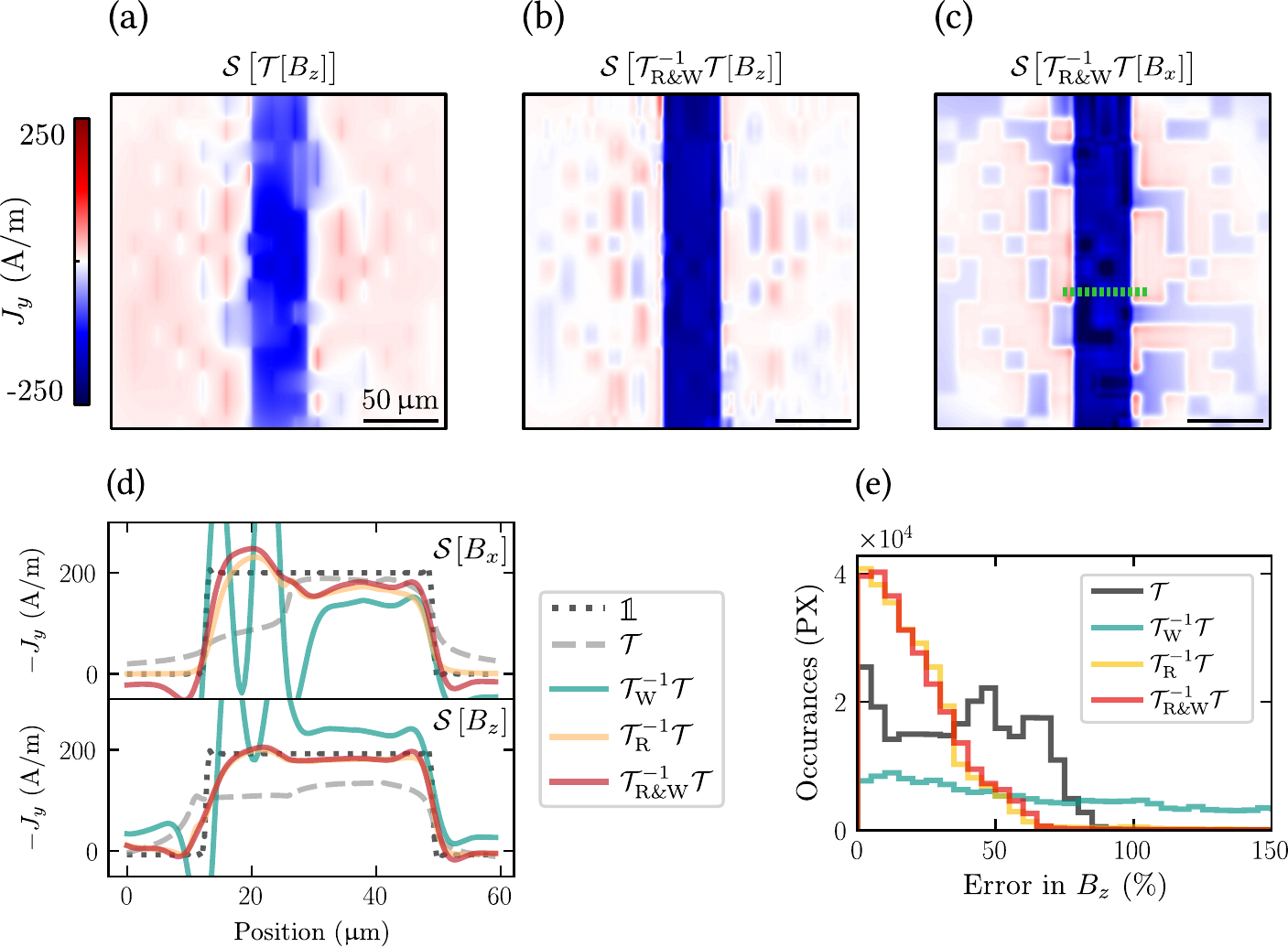}
	\caption{\textbf{Aberration correction on different field components for current simulation.}
	(a)~Current density image ($J_y$ component) for the simulation in Fig.~\ref{fig:current}, with a 1\,$\upmu$m (FWHM) Lorentzian PSF, reconstructed from $\BT_z$ without a deconvolution operation.
	(b)~As in (a), but with an R\&W-deconvolution operation (see text) applied before current reconstruction.
	(c)~As in (b), but with the current reconstruction algorithm utilizing the in-plane component of the magnetic field $B_x$.
	(d)~Line cuts at position highlighted with green in (c) for $J_y$ reconstructed from $B_x$ (top) and $B_z$ (bottom) for various deconvolution methods: aberration-free (black dotted), apparent (grey dashed), W-deconvolved (green), R-deconvolved (orange) and R\&W-deconvolved (red).
	(f)~Histograms of relative error in $B_z$ for each of the deconvolution methods as described above, compared with the apparent case (no deconvolution operation; black line).}
	\label{fig:curr_deconv}
\end{figure*}

Having described the forward problem ($\mathcal{T}$), we now turn to the required backward transform ($\mathcal{T}^{-1}$; Eq.~\ref{eq:Bdeconv}) to correct for these aberrations, which is a deconvolution problem.
We will test the process on the most anomalous images from the previous section, the images of a micromagnet (Sec.~\ref{subsec:mag_deconv}) and a wire carrying current (Sec.~\ref{subsec:curr_deconv}), under a 1\,$\upmu$m (FWHM) Lorentzian PSF with a Gaussian-chessboard emission pattern.

Deconvolution (or image restoration) is a difficult problem in general and an entire discipline in itself.
The goal of this section is thus not to optimize the inversion algorithm in full generality, which should be done for each individual data set once the conditions (e.g.\ PSF, noise) are known.
Instead we aim to validate the feasibility of such a process and identify the key challenges.
To this end we restrict ourselves to only the most basic methods: Wiener-Hunt deconvolution\cite{huntDeconvolutionLinearSystems1972} (non-iterative, denoted `W'), and  Richardson-Lucy deconvolution\cite{richardsonBayesianBasedIterativeMethod1972,lucyIterativeTechniqueRectification1974} (iterative, denoted `R'), both implemented by scikit-image in Python.\cite{waltScikitimageImageProcessing2014}
Each method requires the choice of a suitable regularization parameter; for the simulated data sets shown here we used a balance of unity (Wiener-Hunt), and 15 iterations (Richardson-Lucy).
Additionally, our backward problem actually requires two deconvolution operations (see Eq.~\ref{eq:Bdeconv}), one directly on the PL map $\IT$ and another on the field-PL product $\BT \IT$.
We find that using the Richardson-Lucy method on the former and the Wiener-Hunt method for the latter produces the best results (denoted `R\&W') for simulated data, as will be detailed below.
In contrast, we performed a trial optimisation on the experimental measurements in Sec.~\ref{sec:anom} and decided upon the Richardson-Lucy method with 75 iterations.

The quantitative nature of widefield NV microscopy is often utilized to reconstruct source quantities, such as magnetization or charge current density.\cite{broadwayImprovedCurrentDensity2020,scholtenWidefieldQuantumMicroscopy2021}
We will thus use both the stray magnetic field and the source quantity to estimate our aberration correction fidelity.
For brevity we will not detail the reconstruction process here (which follows the Fourier inversion method in Ref.~\onlinecite{broadwayImprovedCurrentDensity2020}), referring to the operation with the symbol $\mathcal{S}$.
The operation $\mathcal{S}$ will transform the magnetic field ($B_z$ or $B_x$) into a source magnetization $M_z$ and a current density $J_y$ for Sec.~\ref{subsec:mag_deconv} and Sec.~\ref{subsec:curr_deconv}, respectively.

\subsection{Aberration correction: Micromagnet simulation} \label{subsec:mag_deconv}

Our first test simulation is the micromagnet as presented in Fig.~\ref{fig:forward}(a).
As we have seen in the previous section, the apparent magnetic field map imaged through an aberrative optical system [Fig.~\ref{fig:mag_deconv}(a)] is distinctly anomalous where the local PL emission is reduced, alongside a globally reduced magnitude.
Applying the deconvolution operation (R\&W method) now [Fig.~\ref{fig:mag_deconv}(b)], we see a decreased field modulation under low PL squares as well as restored field strengths at the edge of the flake.
The magnetization reconstructions are shown in Fig.~\ref{fig:mag_deconv}(c,d) for Fig.~\ref{fig:mag_deconv}(a,b) respectively.
The magnetization map after the aberration correction has been applied [Fig.~\ref{fig:mag_deconv}(d)] is noticeably larger in magnitude, and with sharper edges.
To quantify the fidelity of the backward process we take line cuts [Fig.~\ref{fig:mag_deconv}(e)] across the flake edge and under a low PL square, for both the magnetic field and magnetization maps.
The Wiener-Hunt method clearly overestimates both the field and the magnetization (green line), while the Richardson-Lucy method appears poorly regulated and is wildly different from the aberration-free profile (black dotted line).
When the two methods are combined  however (red line), both the magnetization and stray field match well, except at the highest spatial frequencies (i.e. there is still significant blurring in these highly quenched regions).
We note that the quantity of interest in such measurements is the areal magnetization inside the flake, where the aberration correction fidelity is relatively good, and sharpness issues at the edge of the flake (or at PL intensity steps) are generally of less importance.
Compared with a reconstruction directly from the aberrant data set (grey dashed line), the combined R\&W method is an order of magnitude improvement in relative error percentage.
Note also that the correction to the magnetization map is roughly a factor of two on the apparent image, aligning with that used in Fu et al.\cite{fuHighsensitivityMomentMagnetometry2020}.
The aberration control fidelity across each magnetic field image for each method can be quantified via histograms of their relative error [Fig.~\ref{fig:mag_deconv}(f)].
A similar histogram for the relative error in the magnetization is qualitatively very similar (not shown).
If no deconvolution method is employed [black line] the relative error is grouped about 50\% -- again agreeing with Fu et al.\cite{fuHighsensitivityMomentMagnetometry2020}.
The Wiener-Hunt method (green line) shifts the weight of the histogram to larger errors, whilst the Richardson-Lucy algorithm (orange line) provides a larger peak at errors below 25\%, with the side effect of a noticeable increase at errors over 100\%.
The combined R\&W method in comparison is much more accurate, shifting the weight of the histogram to a large peak below 10\%, whilst also reducing the weight at the larger errors.

\subsection{Aberration correction: Charge current simulation} \label{subsec:curr_deconv}

Aberration-induced artefacts can be of drastically different magnitudes for different components of a vector magnetic field image, as can be seen in Fig.~\ref{fig:current}(h).
To analyse the influence of this dependence we will now investigate the fidelity of the backward operation on the simulated current density data set, which has more in-plane signal than the micromagnet data set.
We again utilize a 1\,$\upmu$m (FWHM) Lorentzian PSF and a Gaussian-chessboard emission pattern as in Fig.~\ref{fig:current}, however we now reconstruct the source current density, first without a deconvolution operation [Fig.~\ref{fig:curr_deconv}(a)].
A clear modulation of the current density can be seen, as well as a decrease in global current density magnitude (30\%).
We can insert a R\&W-deconvolution operation [Fig.~\ref{fig:curr_deconv}(b)] which is far more accurate, albeit without completely homogenizing the background.
The same process, but using the in-plane component $B_x$  [Fig.~\ref{fig:curr_deconv}(c)] yields an even more anomalous background and clear modulations of the current density in the wire.
Line cuts across the wire [Fig.~\ref{fig:curr_deconv}(d)] highlight that whilst reconstruction from $B_x$ is more accurate away from the emission-quenched regions, these aberrant regions are more complex to correct, and thus  $B_z$ (appropriately corrected) is a preferred reconstruction pathway -- contrary to the aberration-free case where using the in-plane field is advantageous.\cite{broadwayImprovedCurrentDensity2020}
Numeric instability or other forms of inaccuracy at sharp edges appear to dominate the errors, but over the centre of the wire the R and R\&W methods are largely accurate within 10\%.
Counter to the micromagnet case [Fig.~\ref{fig:mag_deconv}(d)] the R and R\&W methods have minimal differences.
This similarity is replicated in a $B_z$ relative error histogram [Fig.~\ref{fig:curr_deconv}(e)], where both the R and R\&W methods clearly outperform the deconvolution-free and W methods.

\subsection{Discussion}

\begin{figure*}[t!]
	\includegraphics[width=0.95\textwidth]{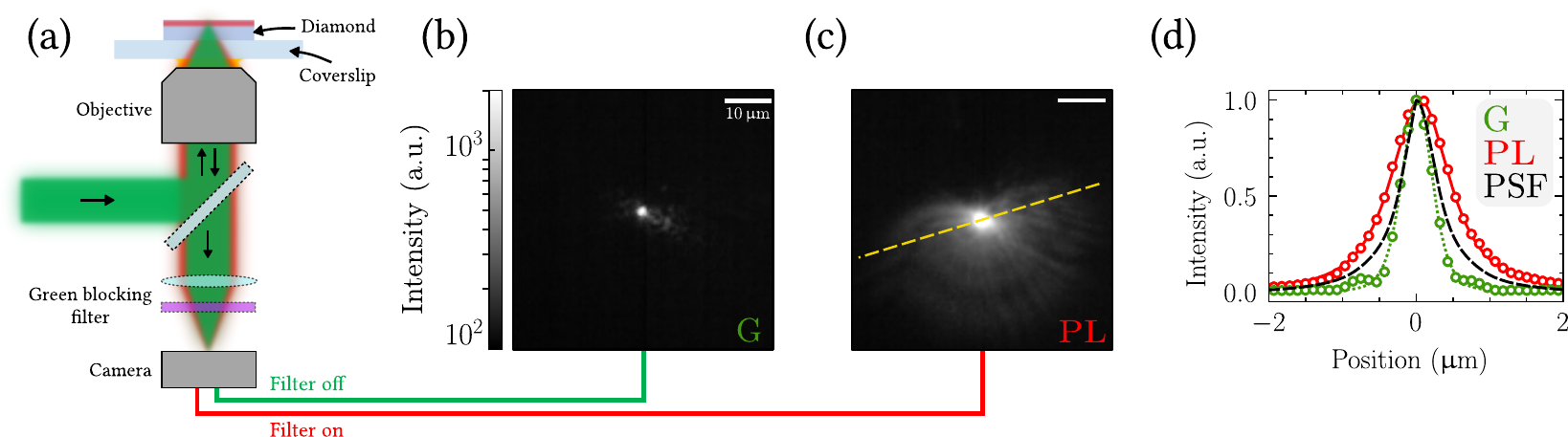}
	\caption{\textbf{Empirical estimation of the point spread function in a quantum diamond microscope.}
	(a)~Experimental set-up used to estimate the PSF. 
	A green laser is focused on the NV layer (which is confined within $\sim10$\,nm of the top surface) and an image is formed on the camera. (b)~Image of the reflected green light (no filter).
	(c)~Image of the PL emission (green light blocked by a filter).
	The images in (b,c) are plot on a logarithmic intensity scale. 
	The asymmetry in the spot shape is attributed to imperfections in the green laser beam. 
	(d)~Line cuts through the green and PL spots, taken along the dashed line shown in (c). 
	The points are the data which are fit to a Moffat distribution (green and red lines for green and PL spots, respectively).
	The dashed black line is the PSF obtained by deconvolution.}
	\label{fig:PSF}
\end{figure*}

We have explored a few deconvolution methods here, but undoubtedly there are other methods that will be better in some or all scenarios.
For example, we determined the Richardson-Lucy method (75 iterations) was the most stable for the experimental measurements in the next section (Figs.~\ref{fig:PSF},\ref{fig:ramp}).
Further work is required to investigate other approaches to improve the fidelity of this procedure, key to accurate quantitative imaging.
We highlight that our work did not simulate the stability of these algorithms to noise in the data, which is an important challenge to the accuracy of aberration-corrected quantitative widefield NV magnetometry.

\section{Resolving the current anomaly}
\label{sec:anom}

We now use our model to re-analyse previously published data from Tetienne et al.\cite{tetienneApparentDelocalizationCurrent2019} that showed an apparent magnetic anomaly. 
The experiments in Ref.~\onlinecite{tetienneApparentDelocalizationCurrent2019} involved imaging the \O rsted magnetic field from current-carrying metallic wires fabricated on a diamond sensing chip, thus realising the situation depicted in Fig.~\ref{fig:current}(a), but the effect of optical aberrations had been ignored. 
Here, we start by estimating the PSF in an imaging set-up similar to that used in Ref.~\onlinecite{tetienneApparentDelocalizationCurrent2019}, and then apply the resulting transformations ${\cal T}$ and ${\cal T}^{-1}$ to the data.

\subsection{Estimation of the PSF}

A common method to estimate the PSF of an optical system is to image a pseudo point emitter.
However, here we want to characterise the PSF in situ, i.e.\ using a dense NV layer near the diamond surface, in the same conditions as for widefield NV imaging. 
To this end, we excite a small area of the NV layer and compare the spot size in the PL image with that of the excited region. 
The experimental set-up is identical to that used for quantum microscopy experiments except that the laser is here tightly focused on the NV layer instead of being defocused [Fig.~\ref{fig:PSF}(a)].

Similar to the conditions in Ref.~\onlinecite{tetienneApparentDelocalizationCurrent2019}, we used a diamond plate of thickness $\approx\,150\,\upmu$m containing a thin layer of NV centres confined within $\sim\,10$\,nm of the surface. 
The diamond plate was glued on a glass coverslip, with the NV layer facing the opposite (air) side. 
Our widefield microscope employs a green laser ($\lambda_{\rm G}=532$\,nm) for excitation of the NV layer in an epi-illumination configuration, and an oil-immersion objective lens (Nikon CFI Super Fluor 40x, ${\rm NA}=1.3$).  
A tube lens (focal length $f=300$~mm) is used to form an image on the camera, either of the NV PL when a bandpass filter is inserted ($\lambda_{\rm PL}=650-750$\,nm) or of the green light reflected off the diamond surface when the filter is removed. 

The green and PL images thus obtained are shown in Fig.~\ref{fig:PSF}(b) and Fig.~\ref{fig:PSF}(c), respectively, plot with a logarithmic intensity scale to emphasise the tails of the spots, while line cuts plot on a linear scale are shown in Fig.~\ref{fig:PSF}(d). 
We verified that the green and PL focal planes coincide, i.e.\ both spots appear smallest for the same position of the diamond along the optical axis. We also made sure the NV PL was not saturated.   

For a perfect optical system free of aberrations, and assuming the rear aperture of the objective is uniformly illuminated, the laser spot in the focal plane (i.e.\ at the NV layer) should be an Airy disk with a FWHM $w_{\rm G}=0.51\lambda_{\rm G}/{\rm NA}\approx210$\,nm; the diffraction limit. 
In our experiments, the laser does not fill the objective uniformly, and additionally we expect aberrations due to the high refractive index of diamond (the objective is designed for imaging through glass). 
Likewise, a point source emitting at $\lambda_{\rm PL}=700$\,nm will ideally form an image on the camera that is an Airy disk with FWHM $w_{\rm PL}=0.51\lambda_{\rm PL}/{\rm NA}\approx270$~nm; this is the diffraction-limited PSF of the imaging system (approximated to a 300-nm Gaussian PSF in Sec.~\ref{sec:sims}). 
But aberrations are also expected due to imaging through the diamond as well as other optical  imperfections. 
These may include intrinsic aberrations from the objective and tube lens, imperfect positioning of the tube lens, and diffuse light scattering within the various optical media (diamond, glue, glass coverslip, immersion oil) and at the interfaces (including at the diamond surfaces which exhibit some roughness).   

We found that the line cuts in Fig.~\ref{fig:PSF}(d) are best fit with a Moffat distribution
\begin{eqnarray}\label{eq:moffat}
    I(r)=I(0)\left[1+\left(\frac{r}{r_0}\right)^2\right]^{-\beta}
\end{eqnarray} 
where $r$ is the distance to the centre, $r_0$ characterises the width of the distribution, and $\beta$ the prominence of the tails ($\beta=1$ corresponds to a Lorentzian distribution while the limit $\beta\rightarrow\infty$ corresponds to a Gaussian distribution). 
The FWHM is given by $w=2r_0\sqrt{2^{1/\beta}-1}$. 
The Mossfat distribution is commonly employed to model the PSF in astronomical imaging, capturing the effects of atmospheric turbulence and diffuse scattering.\cite{anisimovaPointSpreadFunction2013,sandinInfluenceDiffuseScattered2014} 
For the green spot in Fig.~\ref{fig:PSF}(d), we find a mainly Gaussian character ($\beta\approx\,3.3$), albeit wider than expected, FWHM\,$\approx$\,480\,nm.
The NV PL spot is much wider (FWHM\,$\approx$\,940\,nm) and has a Lorentzian character ($\beta\approx\,1.4$), which can be seen by the long tails visible in a log-scale image [Fig.~\ref{fig:PSF}(c)].
Assuming the image of the reflected green light is roughly representative of the actual excitation spot at the NV layer, we can then obtain the PSF $p$ by deconvolving the relation $I_{\rm PL}=\{I_{\rm G} \conv p\}$ where $I_{\rm PL}$ and $I_{\rm G}$ are the measured (and normalised) PL and green distributions.
All deconvolutions in this section are achieved with 75 iterations of the Richardson-Lucy algorithm.
The resulting PSF is shown as a dashed line in Fig.~\ref{fig:PSF}(d), and is best fit with a Moffat distribution with $\beta\approx 1.1$ and FWHM\,$\approx610$\,nm. 
This fit represents a lower bound of the PSF width as the green spot at the NV layer may be smaller than it appears on the image which also includes a convolution. 
We leave a more thorough measurement of the PSF and a detailed analysis of its origin for future work. 

Summarising, we found that in our set-up the PSF is wider than the diffraction limit by a factor of $\sim\,2$ and has a mainly Lorentzian character, with significantly wider tails compared to an ideal Airy pattern (or its Gaussian approximation). 
As we saw in Sec.~\ref{sec:sims}, such tails are an important source of imaging artefacts in spectral imaging.  

\subsection{Aberration prediction and correction}    

\begin{figure*}[t!]
	\includegraphics[width=0.95  \textwidth]{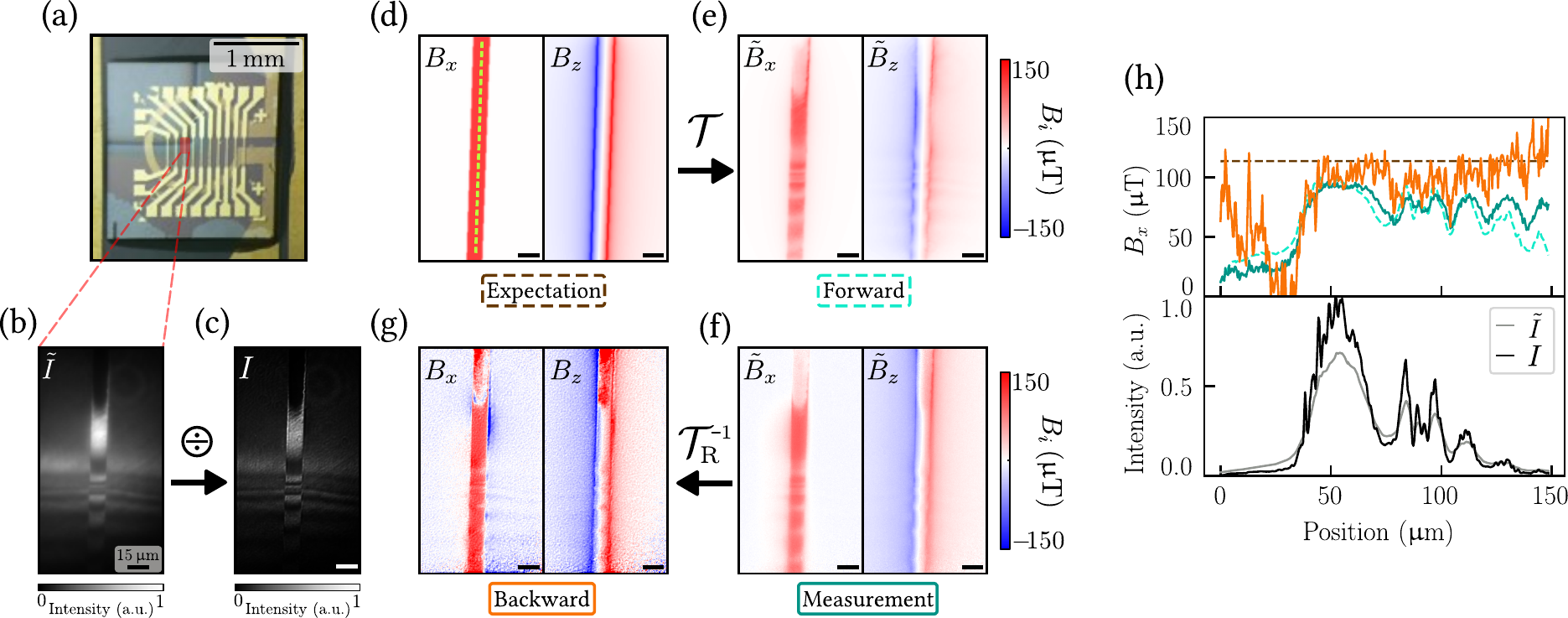}
	\caption{\textbf{Applying aberration model to experimental data set.} 
	(a)~Photo of the NV-diamond sensing chip with current-carrying metallic wires fabricated on top. 
	Before metallisation an AlO$_x$ layer was deposited on parts of the diamond, forming a ramp. 
	(b)~Measured PL image of the NV layer near the AlO$_x$ ramp, with a metallic wire running along the vertical. 
	(c)~True PL emission map obtained by deconvolution of (b) with the PSF estimated in Fig.~\ref{fig:PSF}(d).
	Note (c) is normalized separately to (b).
	(d)~Theoretical magnetic field maps (components $B_x$ and $B_z$) for a current ${\cal I}_c=2$~mA running through the wire, assuming a uniform current density and a standoff $h=60$~nm. 
	(e)~Apparent magnetic field maps ($\BT_x$, $\BT_z$) predicted by applying the forward transformation ${\cal T}$ (Eq.~\ref{eq:B_psf_final}) using (b-d) as inputs.
	(f)~Experimentally measured apparent magnetic field ($\BT_x$, $\BT_z$). 
	(g)~True magnetic field ($\BL_x$, $\BL_z$) reconstructed by applying the backward transformation ${\cal T}_{\rm R}^{-1}$ to (f) via Eq.~\ref{eq:Bdeconv}.
	(h)~Line cuts taken along the wire [dashed line in (d)] showing the $B_x$ field component (top graph) and PL intensity (bottom). The different line cuts are coded (color and solid vs dashed) as indicated in (d-g).}
	\label{fig:ramp}
\end{figure*}

We now use the PSF estimated above to re-interpret the data in Ref.~\onlinecite{tetienneApparentDelocalizationCurrent2019}. 
We examine a particular experiment where the green illumination was highly non-uniform, see photo of the device in Fig.~\ref{fig:ramp}(a) and apparent PL image ($\IT$) in Fig.~\ref{fig:ramp}(b). 
Here, a ramp of AlO$_x$ was deposited on the diamond surface before fabricating the current-carrying metallic wire. 
The varying distance between the NV layer and the top-most reflective interface (oxide-air or oxide-metal) leads to the observed interference pattern. 
Using the PSF estimated in Fig.~\ref{fig:PSF}(d) (i.e.\ Moffat distribution with $\beta\approx1.1$ and FWHM\,$\approx610$\,nm), the true PL emission map $\IL$ can be reconstructed via direct deconvolution [Fig.~\ref{fig:ramp}(c)].
The true PL emission is noticeably sharper and of higher contrast (with respect to the interference pattern) than the raw image acquired.

When a constant current ${\cal I}_c=2$~mA is passed through the metallic wire, an \O rsted magnetic field is produced. 
The theoretical prediction for the in-plane magnetic field component ($B_x$, perpendicular to the current direction) and out-of-plane component ($B_z$) are shown in Fig.~\ref{fig:ramp}(d), where we assumed a uniform current density across the wire and a uniform standoff distance $h=60$\,nm.
We can now apply the transformation ${\cal T}$ (Eq.~\ref{eq:B_psf_final}) to the true field maps to predict the apparent field maps using the PL map $\IT$ and PSF $p$ as inputs. 
The result is shown in Fig.~\ref{fig:ramp}(e) and can be compared with the experimentally measured apparent magnetic field shown in Fig.~\ref{fig:ramp}(f).
The main experimental features are broadly reproduced by the simulation, namely the overall reduction in field strength compared to theory, and the modulation correlated with the PL intensity. 
Conversely, the measured apparent field can be used to reconstruct the true field by applying the inverse transformation ${\cal T}^{-1}$ (here ${\cal T}_R^{-1}$) via Eq.~\ref{eq:Bdeconv}, Fig.~\ref{fig:ramp}(g), which can be compared with the theoretical field, Fig.~\ref{fig:ramp}(d). 
The expected field strength is largely recovered and the modulation attenuated. 

Comparisons are facilitated by examining line cuts along the wire [Fig.~\ref{fig:ramp}(h)]. 
The corrected field is within $\sim20\%$ of the theoretical field (compared to an error of up to 50\% before correction) except in the top part of the images [left-hand side in Fig.~\ref{fig:ramp}(h)] which will be discussed below.
Despite the simplicity of our model and assumptions, it thus provides a solid basis to explain the key features in our initial measurements, and to dramatically improve the accuracy of the estimated field. 
Refining the model to obtain a better agreement is beyond the scope of this work, but here we outline some possible improvements that could be made in the future.
First, the PSF is unlikely to be the same at every point on the sample, especially between regions covered by metal and/or oxide of varying thickness. 
A point-by-point mapping of the PSF and an extension of the model to account for a non-uniform PSF would certainly improve its accuracy.\cite{berishaIterativeMethodsImage2014,sawchukSpacevariantImageRestoration1974} 
Second, the peak finding estimator Eq.~\ref{eq:Bmean} is a crude approximation of the way the magnetic field is estimated in practice (which is via fitting of the electron spin resonance spectrum), in particular it ignores the fact that the spin contrast is not uniform across the image and may vary independently of PL intensity variations, whereas Eq.~\ref{eq:Bmean} applies a weighting proportional to the PL intensity only.
Numerically modelling the exact data fitting process would provide a more accurate description at the cost of increased computation time. 

\subsection{Discussion} 

We conclude this section with a reflection on the implications of our findings on the apparent magnetic field anomaly reported in Ref.~\onlinecite{tetienneApparentDelocalizationCurrent2019}. 
This apparent anomaly is best exemplified by examining the top part of the images in Fig.~\ref{fig:ramp}(b-g), where the apparently measured field [see $\BT_x$ map in Fig.~\ref{fig:ramp}(f)] is much smaller (by a factor of $\sim10$) than the theoretical field [Fig.~\ref{fig:ramp}(d)] under the metallic wire.
In light of our model, we can now explain this apparent discrepancy. 
This top region happens to correspond to a regime where the true PL intensity ($\IL$) is largely quenched by the metallic wire which is in direct contact with the diamond surface in this region. There, the apparent magnetic field is therefore mainly due to delocalised contributions from remote unquenched regions. 
Since these unquenched regions away from the current experience zero magnetic field, the apparent field in the quenched region is very small, as observed, and is largely unrelated to the theoretical expectation. 
Thus, the apparent anomaly is just that -- an {\it apparent} anomaly caused by an extreme case of spectral imaging aberrations. 
It is also unsurprising that the reconstruction in Fig.~\ref{fig:ramp}(g) largely fails in the quenched region since there is little information about the local field in the data, although a refined model as discussed above may improve the outcome.

As for the data set presented in Fig.~\ref{fig:ramp}, all other measurements and observations reported in Ref.~\onlinecite{tetienneApparentDelocalizationCurrent2019} can be explained by the present model, at least qualitatively. 
It is therefore reasonable to conclude that there was no actual anomaly in the electrical behaviour of the device, just imaging artefacts in the magnetic field measurements; the apparent anomaly is resolved. 
This case study illustrates how a seemingly trivial blurring effect can have significant consequences on quantitative analysis of widefield NV imaging measurements, and highlights the importance of image restoration for future experiments.

\section{Conclusion}

Widefield quantum microscopy is an increasingly used technique to map magnetic and other fields, in particular for the quantitative analysis of magnetic materials and current distributions. 
Despite calibration-free accurate measurement a claimed benefit of the technique,\cite{scholtenWidefieldQuantumMicroscopy2021} the accuracy of quantum diamond microscopy has not been characterised in detail thus far.
Here we have shown that optical aberrations can introduce large systematic errors in the measured image exceeding a simple blurring.
A framework was developed to account for the measurement's spectral dimension, both in predicting anomalous measurements as well as a method to control for aberrations via a deconvolution approach.
The combination of large tails on the PSF with local emission quenching were established as the cause of previously reported imaging artefacts, which informed a prescription to identify them in the future.
Further work is required to develop a method to characterise the (possibly spatially variant) response of an imaging system in situ with the diamond and sample present, as well as inform on the source of aberrations.
Other imaging modalities such as spin relaxometry demand a separate and thorough treatment to identify the magnitude of artefacts in their results.
This work will help researchers to identify and control for aberration in widefield quantum microscopy and related techniques and provides a framework to facilitate further improvements to their accuracy.

\begin{acknowledgments}
We thank N.~Dontschuk, D.~McCloskey, D.~Simpson, G.~Balasubramanian and V.~Jacques for stimulating discussions.
This work was supported by the Australian Research Council (ARC) through grants CE170100012, FT200100073 and DP220100178.
I.O.R. and A.J.H. are supported by an Australian Government Research Training Program Scholarship. 
S.C.S gratefully acknowledges the support of an Ernst and Grace Matthaei scholarship.
\end{acknowledgments}

\bibliography{aberrations}	

\end{document}